\documentclass[aps,pra,twocolumn,showpacs,superscriptaddress,floatfix,amsmath,amssymb]{revtex4}
\usepackage{epsfig}
\usepackage{times}

\newcommand{\trace}{\mathop{\rm Tr}\nolimits}

\newcommand{\diag}{\mathop{\rm Diag}\nolimits}

\newcommand{\id}{\mathrm{\openone}}

\newcommand{\be}{\begin{equation}}
\newcommand{\ee}{\end{equation}}
\newcommand{\bea}{\begin{eqnarray}}
\newcommand{\eea}{\end{eqnarray}}
\newcommand{\beas}{\begin{eqnarray*}}
\newcommand{\eeas}{\end{eqnarray*}}

\begin{document}

\title{When are correlations quantum? -- Verification and 
quantification of entanglement by simple measurements.} 

\author{K.M.R.\ Audenaert and M.B.\ Plenio}
\affiliation{Institute for Mathematical Sciences, Imperial College
London, 53 Prince's Gate, London SW7 2PG, UK}

\affiliation{QOLS, Blackett Laboratory, Imperial College London,
Prince Consort Road, London SW7 2BW, UK}

\date{\today}

\begin{abstract}
The verification and quantification of experimentally 
created entanglement by simple measurements, especially 
between distant particles, is an important basic task 
in quantum processing. When composite systems are subjected 
to local measurements the measurement data will exhibit 
correlations, whether these systems are classical or quantum. 
Therefore, the observation of correlations in the classical 
measurement record does not automatically imply the presence 
of quantum correlations in the system under investigation. 
In this work we explore the question of when correlations, 
or other measurement data, are sufficient to guarantee the 
existence of a certain amount of quantum correlations in 
the system and  when additional information, such as the 
degree of purity of the system, is needed to do so. Various 
measurement settings are discussed, both numerically and 
analytically. Exact results and lower bounds on the least 
entanglement consistent with the observations are presented. 
The approach is suitable both for the bi-partite and
the multi-partite setting.
\end{abstract}

\pacs{03.67.Hk,03.65.Ud}

\maketitle

\section{Introduction}
The theoretical and experimental exploration
of entanglement and in particular its characterization,
verification, manipulation and quantification are key concerns of
quantum information science \cite{Plenio V 06}. The resource
character of entanglement is most clearly revealed when dealing
with situations in which a locality constraint is imposed, i.e.
when distributing the state in such a way that subsequent quantum
operations can only act on individual constituents supported by
classical communication. This does not only impose constraints on
the manipulation and exploitation of entanglement but also on its
verification.

In any experiment we will aim to verify the presence of
entanglement by taking measurements. These measurements may either serve
to reconstruct the entire state or may only collect partial
information that is sufficient to reveal the desired entanglement
properties \cite{Sancho H 00,Horodecki 02,Horodecki 01,Horodecki E 02}.
Given that a fundamental goal
in quantum information science is the creation of entanglement
between spatially separate locations one is often forced to assume
that these verification measurements are local as well.
Generically in such verification experiments we will observe
correlations in the measurement record. It is then a natural
question whether these correlations originate from quantum
correlations in the underlying state or can be explained by
a classically correlated separable state. Then, if there are
quantum correlations, one can ask how much
quantum correlations are guaranteed to be there, given the measurement data.

Consider as an example a two-qubit system and the measurement of
correlations between Pauli-operators along the z-axis, i.e.\ the
quantity
\begin{equation}
    \label{example1}
    {\cal C}_{zz}(\rho_{AB}) =
    \trace[\rho_{AB}(\sigma_z\otimes\sigma_z)]
    - \trace[\rho_{A}\sigma_z]\trace[\rho_{B}\sigma_z],
\end{equation}
where $\rho_A$($\rho_B$) are the reduced density operators
resulting from tracing out party B (A) in the original state
$\rho_{AB}$. If ${\cal C}_{zz}(\rho_{AB})=-1$ then the measurement
outcomes are perfectly anti-correlated and are thus exhibiting
very strong, albeit negative, correlations. Do such correlations
imply the existence of quantum correlations in the underlying
quantum state? To decide this we must address the following

{\em {\bf\em Fundamental Question}: What is the entanglement
content of the least entangled quantum state that is compatible
with the available measurement data?}

Mathematically, this question is formulated as a minimization
problem in which the amount of entanglement in the underlying
quantum state must be minimized subject to the constraints
imposed by the measurement data as well as by the positivity
and normalization of the state \cite{footnote2}. The measurement
data will be the expectation values of some observables $A_i$
or some non-linear function $F_i(\rho)$ of the density matrix. Then
the minimal amount of entanglement $E_{\min}$ under the given
constraints is given by
\begin{eqnarray}
    E_{\min} = \min_\rho\{
    E(\rho): \trace[\rho A_i]=a_i, F_i(\rho)=f_i
    \}
    \label{minimization}
\end{eqnarray}
where the minimisation domain is the set of states $\rho$
and $E(\rho)$ is the entanglement measure of
choice \cite{Plenio V 06}. Note that this formulation
applies equally to the bi-partite as to the multi-partite
setting. Note that the importance of the minimization of entanglement
in quantum state reconstruction in quantum information theory was
also pointed out in the context of Jaynes' principle
\cite{Horodecki HH 99}.

The mathematical minimization problem formulated by eq.\ (\ref{minimization})
may be addressed by
techniques from optimization theory (see e.g.\ \cite{Boyd V 05}).
If the constraint are all linear and the entanglement quantifier
is convex then methods from convex optimization theory may be
applied. More complicated constraints that are not linear in
the density operator (e.g.\ purity measures) can complicate matters
considerably. Generally it will not be possible to obtain analytic
solutions to the optimization problem and techniques to obtain
lower bounds or numerical approaches must be used. The analytical and
numerical exploration of these issues will be the main purpose
of this work.

If the optimal state in eq.\ (\ref{minimization}) is separable,
i.e.\ $E_{\min}=0$, then in
reply to our fundamental question we must conclude that the
available correlations in the measurement record do not imply
quantum correlations in the underlying quantum state.
It might be, but need not be entangled. Indeed, in
the example given in eq.\ (\ref{example1}), the least entangled
state compatible with the observation ${\cal
C}_{zz}(\rho_{AB})=-1$ is given by
\begin{equation}
    \rho = \frac{1}{2}\left(|01\rangle\langle 01| + |10\rangle\langle 10|\right)
\end{equation}
which is clearly a separable state. Therefore, the observation of
classical correlations for the measurement along one set of
directions alone is not sufficient for the verification of
entanglement. This well-known observation in quantum
information science is particularly relevant in experimental
situations where only a very restricted set of measurement
settings is available.

One way forward consists in measuring additional
observables. For example, one may consider the measurement of
\begin{equation}
    {\cal C}_{{\vec n}{\vec n}}(\rho_{AB}) =
    \trace[\rho_{AB}\sigma_{{\vec n}}\otimes\sigma_{{\vec n}}]
    - \trace[\rho_{A}\sigma_{{\vec n}}]\trace[\rho_{B}\sigma_{{\vec n}}]
\end{equation}
for all spatial directions ${{\vec n}}$. Observation of perfect anti-correlations in
all of these measurement records then uniquely identifies the
singlet state
   $ |\psi\rangle = \frac{1}{\sqrt{2}}(|01\rangle-|10\rangle)$
 as the only state compatible with all such measurements. This
state carries one ebit of entanglement.

In other experimental situations it may be possible to assert that
the state possesses a certain minimal degree of purity \cite{Purity}, e.g.\ when
decoherence rates, or at least upper bounds for it, are known. Let us for
example assume that we know not only that ${\cal
C}_{zz}(\rho_{AB})=-1$ but also that $\trace[\rho_{AB}^2]=1$, i.e.\ that
the underlying quantum state is pure. Then again it is
straightforward to conclude that the only states compatible with
these two assumptions are of the form
\begin{equation}
    |\psi\rangle = \frac{1}{\sqrt{2}}(|01\rangle +
    e^{i\phi}|10\rangle),
\end{equation}
that is, quantum states with one ebit of entanglement.

These simple examples serve to make two points. Firstly, the simple observation of
correlations in measurements along a single fixed orientation is not
enough to guarantee entanglement in the underlying quantum state.
Secondly, additional information, be it correlation
measurements along different directions or information about the purity
of the states, may be sufficient to ensure that the correlations
found in the classical measurement record indeed prove
entanglement in the underlying quantum state.
Needless to say, in general
the situation is quite involved as the measurement data may be more varied
than those in the above examples. It should also be noted that the
local measurement of the correlation
functions mentioned above often implies that we possess more
information than just these correlations. Indeed, we will often
possess local statistics as well, which in turn can be taken into
account when answering our fundamental question concerning the
minimal entanglement compatible with the measurement data.
Generally, when we are provided with an entangled state, then
any additional information will make it less and less likely
that the measurement data is compatible with a separable state.

Our fundamental question is of particular relevance in experimental
settings in which it is difficult to perform measurements for an
arbitrarily large number of measurement settings, as is required for doing full state
tomography. This may be the
case for example in solid state physics, where it is not
always straightforward to perform arbitrary measurements. Another
reason may simply be the existence of constraints on the measurement time,
dictated for example by the stability time of an experiment
(e.g.\ in interferometric setups in optics)
or by the decoherence time (in solid state or other
implementations).

The present work shares some relations with \cite{Guehne RW 06},
\cite{Eisert BA 06} and \cite{Cavalcanti T 06} where similar questions are
developed but where emphasis is placed on observables that are obtained from
the theory of entanglement witnesses \cite{Witnesses}. Other approaches
are considered in \cite{Mintert B 06,Carteret 05,curty,wolf,toth}. While
these, as well as the present work, consider the analysis of a
specific state, a somewhat different approach is taken in \cite{Dynamical}.
Here the dynamics of the gate used to produce entanglement is considered
while measurements are restricted to a single measurement basis. The
approach is to make repeated measurements during the gate's time evolution.
This contrasts with our approach which only requires to make measurements on the
final state, irrespective of the process that created it.

In this paper we will address our fundamental question for systems consisting only of
qubits, as this is by far the most relevant system from an experimental
viewpoint. I should be noted however that the approach remains valid unchanged
for qudits or even infinite dimensional systems.
We begin with an illustration of the general approach in
which correlations and purity are quantified by quantum mutual
information and global entropy, respectively. While these quantities are not directly
experimentally measurable, they allow for the fundamental question to be most easily answered.
Then we consider the question for
correlations between measurements of Pauli-operators along a
single axis, e.g.\ the z-axis. In the process we prove an
inequality between correlations and purity that is valid exactly
if a two-qubit state is separable and use it to provide necessary
and sufficient conditions for entanglement to be inferred.
Subsequently, we consider correlations along two different
measurement axes, e.g.\ x-x correlations and z-z correlations.
Finally we consider the situation in which we take into account
the local expectation values that are obtained in most experiments
to sharpen the verification of entanglement. We finish with some
conclusions.

\section{Mutual Information, Entropy and Entanglement}
To illustrate the general approach that we are advocating, we begin by considering
a situation in which the known system properties are
the entropy of the state (determining the state's purity) and
the quantum mutual information (determining the state's correlation), and
in which the entanglement measure of choice is the relative entropy of entanglement
\cite{Adesso SI 03}.
The reason for this choice is that there exists a very simple relationship between these quantities,
and the solution of the minimization problem eq.\ (\ref{minimization}) is immediate.

The quantum mutual information is given by
\begin{equation}
    \label{mutualinformation}
    I(\rho_{AB}) = S(\rho_A) + S(\rho_B) - S(\rho_{AB}),
\end{equation}
and the relative entropy of entanglement is \cite{Vedral PRK 97,Vedral P 98,
Plenio V 06}
\be
E_R(\rho) = \min_\sigma\{S(\rho||\sigma): \sigma\mbox{ separable}\}.
\ee

\begin{figure}[tbh]
\centerline{
\includegraphics[width=8.4cm]{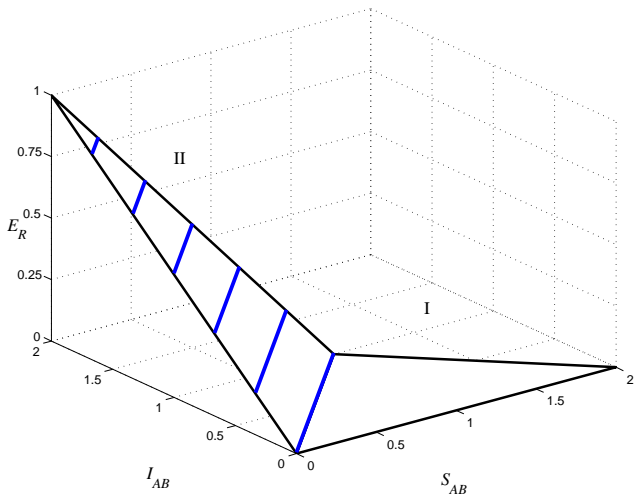}}
\caption{\label{figp1} State space in the $S_{AB}$-$I_{AB}$ plane,
depicting the minimal possible entanglement $E_R$ for every point.
In region I there is no guaranteed entanglement, while in region II
one has at least $E_R\ge (I_{AB}-S_{AB})/2$.}
\end{figure}

For a 2-qubit state, the physically possible values of the pair $(I_{AB},S_{AB})$
are located in a triangle spanned by the points $(0,0)$, $(2,0)$ and $(0,2)$ (see Figure \ref{figp1}).
That is, $I_{AB}\ge0$, $S_{AB}\ge0$ and $I_{AB}+S_{AB}\le 2$; equality in the latter inequality is obtained
when both reductions $\rho_A$ and $\rho_B$ are maximally mixed.

The solution to eq.\ (\ref{minimization}) is obtained by applying an inequality lower bounding the relative entropy of
entanglement \cite{Plenio VP 00} and showing that equality can be achieved for every pair of values of $(I_{AB},S_{AB})$.
The inequality is
\begin{equation}
    E_R(\rho_{AB}) \ge \max\{S(\rho_A)-S(\rho_{AB}),S(\rho_B)-S(\rho_{AB})\},
\end{equation}
which directly implies \cite{footnote3}
\begin{equation}
    \label{bound}
    E_R(\rho_{AB}) \ge \max(0,\frac{1}{2}(I(\rho_{AB})-S(\rho_{AB}))).
\end{equation}
The bound is zero in region I ($S_{AB}\ge I_{AB}$) and non-zero in region II ($S_{AB}< I_{AB}$).
Equality in region I is obtained by diagonal states; they cover region I completely, and as any diagonal state
is separable, they have $E_R=0$.
Equality in region II is obtained by so-called maximally correlated states, which are of the form
\begin{equation}
    \label{minimal}
    \rho_{AB} = a|00\rangle\langle 00| + b|00\rangle\langle 11|
    + b |11\rangle\langle 00| + (1-a)|11\rangle\langle 11|.
\end{equation}
Example 3 of \cite{Vedral P 98} shows that
these states satisfy
\begin{equation}
    E_R(\rho_{AB}) = S(\rho_{A}) - S(\rho_{AB}) = S(\rho_{B}) - S(\rho_{AB}).
    \label{value}
\end{equation}
For any given value of $I_{AB}$ and $S_{AB}$ in region II
we can find a state of the form eq.\ (\ref{minimal}) realizing these values. By
eq.\ (\ref{bound}) and eq.\ (\ref{value}) this state realizes the
smallest possible value for the $E_R$ given $I_{AB}$ and $S_{AB}$.

The upshot of the results obtained here is that knowledge of the two quantities $S_{AB}$
and $I_{AB}$ allows one to have much better bounds on the entanglement $E_R$ than with
just knowledge of the correlations $I_{AB}$ alone.
Indeed, without knowing the purity $S_{AB}$, one has to assume the worst case, being
$S_{AB}=2-I_{AB}$, in which case the lower bound on $E_R$ is given by
$$
E_R\ge\left\{
\begin{array}{ll}
0,& I_{AB}\le1 \\
I_{AB}-1,& I_{AB}\ge1.
\end{array}
\right.
$$
If, on the other hand, the state is known to be pure, say, ($S_{AB}=0$)
then the much sharper bound
$$
E_R\ge I_{AB}/2
$$
can be obtained.

In the rest of the paper we will apply the approach illustrated here for studying the main question
eq.\ (\ref{minimization}) in the context of experimentally accessible quantities.
In the next Section the measure of correlation will be based on measurements along the z-axis. It will turn out that
without knowledge of the purity one cannot find any lower bound on entanglement other than the trivial bound $E\ge0$.
Thus, while in the present Section one can get some information about the entanglement from the quantum mutual information
without knowledge of the purity, in the next Section knowledge of the purity is absolutely essential.
\section{Purity and Correlations}\label{sec:purity}
In a number of experimental
settings it is not straightforward to carry out measurements
along arbitrary directions. To obtain a measure of correlation in those settings, one
can for example consider the quantity
\begin{equation}
    \label{correlations}
    {\cal C}_{zz}(\rho_{AB}) =
    \trace[\rho_{AB}(\sigma_z\otimes\sigma_z)]
    - \trace[\rho_{A}\sigma_z]\trace[\rho_{B}\sigma_z],
\end{equation}
which only requires measurements along the particles' $z$-axes.
However, in the previous Section we already alluded to the fact that knowledge of this correlation measure
alone is not sufficient to prove the presence of quantum entanglement. We will establish that fact in the present Section.
Moreover, we will show that if in addition the purity of the state is known,
as quantified by
\begin{equation}
    \label{purity}
    {\cal P}(\rho_{AB}) = \frac{4}{3}\left(\trace[\rho_{AB}^2] - \frac{1}{4}\right),
\end{equation}
and provided this purity is large enough,
then and only then can one infer entanglement from the z-correlation measure.

Now the question is: how pure does
the underlying quantum state have to be so that $|{\cal C}_{zz}|>0$
indeed implies quantum entanglement? Or, more precisely:

{\em When are all states consistent with given values of ${\cal C}_{zz}$ and ${\cal P}$
non-separable, and what is the least entanglement compatible with these values?}

It turns out that the rigorous analytical answer is surprisingly
involved, largely due to the non-linearity of the constraints
involved in the minimization problem, especially if one is also
interested in the actual amount of entanglement that can be
guaranteed from such measurements.

As measure of entanglement we have used the logarithmic negativity $E_N$, because this is the
measure that is most easily calculated \cite{footnote}.
The log-negativity is defined as
$$
E_N(\rho_{AB}) := \log\trace|\rho^\Gamma|,
$$
where $\Gamma$ denotes partial transposition w.r.t.\ subsystem B.

\begin{figure}[tbh]
\centerline{
\includegraphics[width=8.4cm]{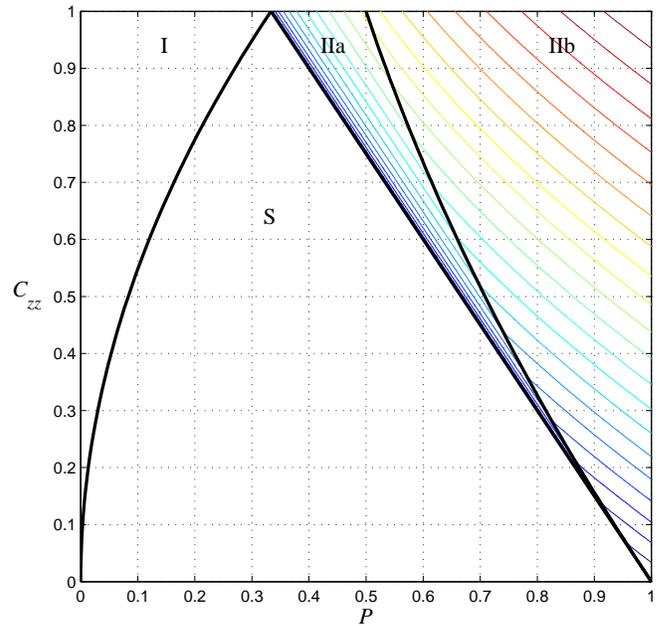}}
\caption{\label{fig0} This plot show the numerical results on the
smallest amount of entanglement that is compatible with given
values of purity (see eq.\ (\ref{purity})) and correlations in the
measurement record (see eq.\ (\ref{correlations})).
The logarithmic negativity $E_N$ is shown as a contour plot in function of the
parameters $\cal P$ and ${\cal C}_{zz}$.
Four regions can be distinguished. In region I no state exists that is compatible
with the specified values of $\cal P$ and ${\cal C}_{zz}$.
A large central
region, denoted by $S$, does not allow to conclude the presence of entanglement.
Regions IIa and IIb are the only ones where all compatible states have
non-zero entanglement $E_N$. For these regions, the minimal $E_N$ is given by
eqns.\ (\ref{eq:IIa}) and (\ref{eq:IIb}), respectively.}
\end{figure}

In Figure \ref{fig0}, we present our numerical results on the smallest amount of
entanglement compatible with given values of purity (see
eq.\ (\ref{purity})) and of correlations in the measurement record (see
eq.\ (\ref{correlations})). This numerical evaluation suggests the following:
\begin{itemize}
\item Region I does not allow for any physical states.
\item There is a well-defined central region $S$ that does not allow to
infer the presence of entanglement as the values for purity and
correlations can be reproduced by a separable state.
\item Only in regions IIa and IIb is entanglement guaranteed.
The minimal value of $E_N$ in those regions is given by
\be\label{eq:IIa}
E_N\ge \log_2(1+\sqrt{2(Q-1)+{\cal C}_{zz}}),
\ee
in Region IIa, and
\be\label{eq:IIb}
E_N\ge\log_2({\cal C}_{zz}+\sqrt{2Q-1})
\ee
in Region IIb.
Here, $Q=\trace[\rho^2]=(3{\cal P}+1)/4$.
\end{itemize}
One may either calculate both bounds and take the minimum, or infer which region one is in via the limits
\be
1-\frac{{\cal C}_{zz}}{2} \le Q \le \frac{1}{2}\left(1+\left(1-\frac{{\cal C}_{zz}}{2}\right)^2\right),
\ee
which hold for Region IIa.

We stress that we do not have a complete proof of these statements.
They were derived -- in a rather laborious way -- starting from an Ansatz concerning the form of the states achieving the bounds.
This Ansatz was in turn obtained from a combination of Monte-Carlo calculations and inspired guess-work.
While a proof does not seem forthcoming, the numerical evidence for correctness of the Ansatz,
and of the bounds derived from them, is very convincing.
The interested reader is advised to contact the authors for further details.

{\em Analytical proof of boundaries --}
What we do have been able to prove is the analytical form of the boundaries of the $S$ region, the region where
be separable states. They are given by
\begin{equation}
    \label{boundaries}
    \frac{{\cal C}_{zz}^2}{3} \le {\cal P} \le 1 - \frac{2 {\cal C}_{zz}}{3}.
\end{equation}
Here, the first inequality defines the boundary with Region I while the second one
defines the boundary with Region IIa.

To proceed, we treat boundary I and II
separately. For boundary I we can simplify the form of $\rho$ that
needs to be considered quite significantly. To this end note that
correlations ${\cal C}_{zz}(\rho)$ are unaffected by the
transformation
\begin{eqnarray*}
    \rho \mapsto {\bar \rho}&=&
    \frac{1}{4}\left(\rho +
    (\id\otimes\sigma_z)\rho(\id\otimes\sigma_z)+
    (\sigma_z\otimes\id)\rho(\sigma_z\otimes\id)\right.\\
    && \left. \;\;\;\; +
    (\sigma_z\otimes\sigma_z)\rho(\sigma_z\otimes\sigma_z)\right)
\end{eqnarray*}
ie, ${\cal C}_{zz}(\rho)={\cal C}_{zz}({\bar \rho})$ but at the
same time the transformation from $\rho$ to ${\bar \rho}$ reduces
purity as they correspond to pinchings \cite{Bhatia}. A state
${\bar \rho}$ that is invariant under the above maps is
diagonal. As these maps are local we find that if $E(\rho)=0$,
then $E({\bar \rho})=0$. To determine the boundary I let us now
fix a value for ${\cal C}_{zz}$ and determine the smallest purity
compatible with it. If we have a $\rho$ with a given purity then
by the above transformations we can find a ${\bar\rho}$ with the
same ${\cal C}_{zz}$ and no larger purity that is diagonal.
Therefore it is sufficient to restrict attention from the outset
to diagonal $\rho$, i.e.\ $\rho = \diag(a,b,c,1-a-b-c)$.
Then we find ${\cal C}_{zz} = 1-2b-2c$ and the purity is given by
\begin{equation}
    {\cal P} = \frac{4}{3}\left(a^2+b^2+c^2+(1-a-b-c)^2-\frac{1}{4}\right).
\end{equation}
Without restriction of generality we assume ${\cal C}_{zz}\ge 0$
(the case ${\cal C}_{zz}\le 0$ can be treated analogously) and one
finds that the purity is minimized for $b=c$. This leaves us with
the minimization of the expression
\begin{equation}
    {\cal P} = \frac{4}{3}\left(a^2+2b^2+(1-a-2b)^2-\frac{1}{4}\right)
\end{equation}
for
\begin{equation}
    {\cal C}_{zz} = 1-4b.
\end{equation}
Then the minimal purity compatible with the given ${\cal C}$ is then found to be
\begin{equation}
    {\cal P} = \frac{{\cal C}_{zz}^2}{3}
\end{equation}
yielding the boundary confirming eq.\ (\ref{boundaries}).

Determining the boundary II is more involved and is based
on the observation that
for all separable states $\rho$ we have \cite{Hall AB 06}
\be\label{th1}
        \trace[\rho^2]+\frac{1}{2} {\cal C}_{zz}(\rho)\le 1.
\ee
We first note that $\trace[\rho^2]+\frac{1}{2} {\cal C}_{zz}(\rho)$
is convex in $\rho$. Indeed, a short calculation reveals that this expression is equal to
$$
\sum_{j\neq k} |\rho_{jk}|^2  + 1+2(\rho_{22}^2-\rho_{22}+\rho_{33}^2-\rho_{33}).
$$
As every term is convex in $\rho$, the total expression is.
Therefore, the inequality only has to be checked for
the extremal points of the set of separable states, i.e.\ for
pure product states. This, however, is very easy: for product states, ${\cal C}_{zz}=0$,
and for pure states $\trace[\rho^2]=1$, whence the inequality is satisfied with equality.

Now we note that the separable states $\rho = a|00\rangle\langle
00|+(1-a)|11\rangle\langle 11|$ saturate the bound (\ref{th1}).
Rewriting this bound in terms of ${\cal P}(\rho)$ we find ${\cal P}\le 1-2{\cal C}_{zz}/3$.
This then completes the proof for boundary II.

{\em A lower bound for $E_N$ --} As mentioned above, we have not been able to prove our lower
bounds (\ref{eq:IIa}) and (\ref{eq:IIb}) so far. Nevertheless, inequality (\ref{th1}) suggests that
\begin{equation}
    E_{lower} = \log_2^+(\trace[\rho^2]+\frac{1}{2} {\cal C}_{zz}(\rho))
    \label{lowerpurity}
\end{equation}
might be a lower bound on the entanglement in all regions.
Here we define the function $\log_2^+(x) := \max(0,\log_2(x))$; that is, $\log_2^+(x) = 0$ for $x\le1$.
We will prove eq.\ (\ref{lowerpurity}) in subsection
\ref{LowerBound}, where a general recipe for the derivation of
such bounds is presented.

%
\section{Correlations along different directions \label{sec:xxzz}}
Let us now move away from the use of non-linear properties of the density
operator such as purities or entropies and consider only linear functionals,
i.e.\ expectation values of quantum mechanical operators that are directly
accessible to experimental detection. Consider the case when we are given the quantities
\beas
C_{zz} &=& \trace[(\sigma_z^{(1)}\otimes\sigma_z^{(2)})\rho] \\
C_{xx} &=& \trace[(\sigma_x^{(1)}\otimes\sigma_x^{(2)})\rho]
\eeas
(note that these are different quantities than the one
used in the previous Section). In this case
it is quite straightforward to determine the minimal entanglement
compatible with any choice of $C_{xx}$ and $C_{zz}$. To see this
we first realize that $C_{xx}$ and $C_{zz}$ are invariant under
the transformation
\begin{equation}\label{cxxczztrf}
    \rho \rightarrow \rho' =
    \frac{1}{4}\sum_{i=0,x,y,z}
    (\sigma_i\otimes\sigma_i)\rho(\sigma_i\otimes\sigma_i) \, .
\end{equation}
Thus for given $C_{xx}$ and $C_{zz}$ we may restrict attention to
states of the form
\begin{equation}\label{cxxczzstates}
    \rho = \left(\begin{array}{cccc}
    \frac{1+C_{zz}}{4} &          0         &      0      & \frac{C_{xx}}{2}-b \\
              0        & \frac{1-C_{zz}}{4} &      b      &        0           \\
              0        &          b         & \frac{1-C_{zz}}{4} & 0           \\
    \frac{C_{xx}}{2}-b &          0         &      0      & \frac{1+C_{zz}}{4}
    \end{array}\right)
\end{equation}
Let us now consider the case $C_{xx}\ge 0$ and $C_{zz}\ge 0$. Any
other choice can be reduced to this one by application of
$\id\otimes \sigma_x$ or $\id\otimes \sigma_z$ onto the state.

The requirements for positivity of $\rho$ are
$b\le (1-C_{zz})/4$ and $C_{xx}/2-b\le (1+C_{zz})/4$.
From the first requirement follows that $|b|\le(1+C_{zz})/4$.
Thus any amount of negativity of the partial transpose
of $\rho$ must arise from $\rho_{14}=C_{xx}/2-b$. As we are looking
for the smallest amount of entanglement compatible with the choice
$C_{xx},C_{zz}\ge 0$, we must minimize
$C_{xx}/2-b$, i.e.\ maximize $b$. This is achieved by the
choice $b=(1-C_{zz})/4$; one checks that this choice satisfies the second requirement $C_{xx}/2-b\le (1+C_{zz})/4$.
Then we find $E_{\min}(C_{xx},C_{zz}) = \log_2^+ (C_{xx}+C_{zz})$. For general $C_{xx},C_{zz}$ we find
\begin{equation}
    E_{\min}(C_{xx},C_{zz}) = \log_2^+ (|C_{xx}| + |C_{zz}|).
    \label{cxxczz}
\end{equation}

This result may easily be generalized to the case of three
correlations
\beas
C_{zz}&=&\trace[(\sigma_z^{(1)}\otimes\sigma_z^{(2)})\rho], \\
C_{xx}&=&\trace[(\sigma_x^{(1)}\otimes\sigma_x^{(2)})\rho], \\
C_{yy}&=&\trace[(\sigma_y^{(1)}\otimes\sigma_y^{(2)})\rho],
\eeas
for which we find
\begin{equation}
    E_{\min} = \log_2^+ ((1+|C_{xx}|+|C_{yy}|+|C_{zz}|)/2). \label{eq27}
\end{equation}

%
\section{Local statistics from correlation measurements improves entanglement
estimation}
If the sub-systems for which we would like to verify 
entanglement are distant, then any measurement strategy 
has to be composed of local measurements. In this way 
we can, of course, still obtain averages such as 
$\langle \sigma_x\otimes\sigma_x\rangle$ by measuring 
local observables (such as $\sigma_x$) and use these 
averages to determine correlations (such as $C_{xx}$).
While the assessment of entanglement wil primarily depend 
on the values of these correlations, it is important
to note that these local measurements will in addition 
yield local averages (such as $\langle \sigma_x\rangle$),
which by themselves are not useful to assess entanglement, 
but when taken together with the correlation values 
represent additional knowledge that we can and should 
take account of. Note that the question of the verification 
of the presence of entanglement in the particular setting 
considered in this section has been addressed in \cite{curty}.
The full analytical treatment of the quantification 
of the least amount of entanglement compatible with the
measurement data in this setting is quite complicated 
due to the large number of possibilities that are 
available. In the following we will simply present an 
example to illuminate the impact that additional local 
information may have on the question of assessing least 
entanglement compatible with the measurement data.

Let us reconsider the case in which we employed
$C_{zz}=\trace[(\sigma_z^{(1)}\otimes\sigma_z^{(2)})\rho]$ 
and $C_{xx}=\trace[(\sigma_x^{(1)}\otimes\sigma_x^{(2)})\rho]$. 
In this setting we found that 
$E_{\min}(C_{xx},C_{zz}) = \log_2^+ (C_{xx}+C_{zz})$ (eq.\ (\ref{cxxczz})).
Let us now investigate what can be gained by taking 
into account knowledge of $z_1:=\trace[\sigma_z^{(1)}\rho]$ and
$z_2:=\trace[\sigma_z^{(2)}\rho]$; that is, we determine the minimal amount of entanglement compatible with
the information given in $C_{xx},C_{zz}$ \textit{and} $z_1,z_2$.

We can no longer restrict ourselves to states of the form (\ref{cxxczzstates}), because $z_1$ and
$z_2$ are not invariant under transformations (\ref{cxxczztrf}).
The optimal states
can now be assumed to possess a $\sigma_z\otimes \sigma_z$ symmetry.
The diagonal elements of the optimal $\rho$ are fully determined by
$C_{zz},z_1,z_2$ and $\trace[\rho]=1$.
Employing the $\sigma_z\otimes \sigma_z$ symmetry of the system the problem can be reduced to a
single-parameter minimisation. The optimal states turn out to be of the form
\be
\rho=\left(
\begin{array}{cccc}
a&0&0&f \\
0&b&e&0 \\
0&e&c&0 \\
f&0&0&d
\end{array}
\right),
\ee
with
\beas
a&=&(1+z_1+z_2+C_{zz})/4 \\
b&=&(1-z_1+z_2-C_{zz})/4 \\
c&=&(1+z_1-z_2-C_{zz})/4 \\
d&=&(1-z_1-z_2+C_{zz})/4 \\
e+f&=&C_{xx}/2,
\eeas
and
\beas
0\le &e&\le\sqrt{bc} \\
0\le &f&\le\sqrt{ad}.
\eeas
Given $z_1$ and $z_2$, there are now restrictions on the values of $C_{xx}$ and $C_{zz}$:
\beas
C_{zz} &\le& 1-|z_2-z_1| \\
C_{xx} &\le& \frac{1}{2}\sqrt{(1+C_{zz})^2-(z_1+z_2)^2} \\
&& +\frac{1}{2}\sqrt{(1-C_{zz})^2-(z_1-z_2)^2}.
\eeas
The negative eigenvalue of the partial transpose of $\rho$ is given by
\beas
\lambda_{\min} &=& \frac{1}{4}\min\big(1+C_{zz} - \sqrt{(z_1+z_2)^2+(4e)^2}, \\
&& \quad 1-C_{zz}-\sqrt{(z_1-z_2)^2+(4f)^2}\big).
\eeas
The log-negativity $E_N$ is then
$$
E_N=\log_2(1-2(\min(0,\lambda_{\min})).
$$
To minimise $E_N$, we have to maximise $\lambda_{\min}$ over all allowed values of $e$,
which is the range
\beas
\max(0,C_{xx}/2-\sqrt{ad}) \le e \le \min(C_{xx}/2,\sqrt{bc}).
\eeas

\begin{figure}[tbh]
\centerline{
\includegraphics[width=8.5cm]{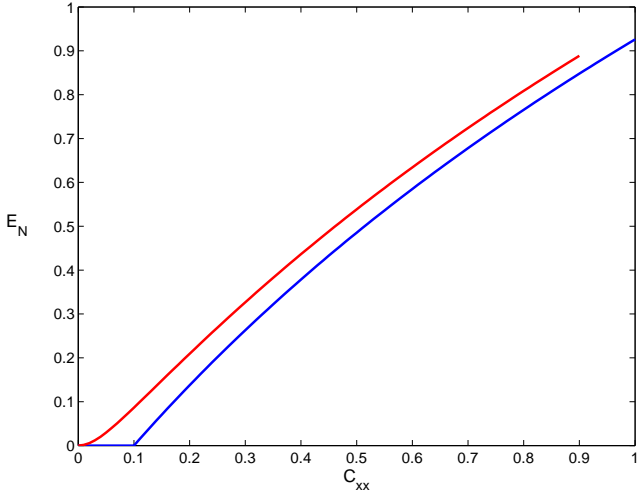}}
\caption{\label{additional} For given $z_1:=\trace[\sigma_z^{(1)}\rho]$
and $z_2=\trace[\sigma_z^{(2)}\rho]$ we plot the difference between
the minimal entanglement (quantified by the logarithmic negativity)
compatible with the observation of
$C_{zz}=\trace[(\sigma_z^{(1)}\otimes\sigma_z^{(2)})\rho]$ and
$C_{xx}=\trace[(\sigma_x^{(1)}\otimes\sigma_x^{(2)})\rho]$ and the
minimal entanglement without the constraints imposed by $z_1$ and $z_2$.
In this example, $C_{zz}=0.9$.
The lower curve is the minimal entanglement without knowledge of $z_1$ and $z_2$ (hence the worst case was assumed);
the upper curve is the minimal entanglement with $z_1=0.3$ and $z_2=0.2$.}
\end{figure}

As an example, in Figure \ref{additional}, we
present the difference between the minimal compatible entanglement for given
$(C_{zz}=1-|z_1-z_2|=0.9,C_{xx},z_1=0.3,z_2=0.2)$ and the one when only $(C_{zz}=0.9,C_{xx})$ are given.
For the given value of $C_{zz}$, either $b=0$ or $c=0$, so that the only allowed value for $e$ is $e=0$, giving
$$
\lambda_{\min} = \frac{1}{4}(1-C_{zz}-\sqrt{(z_1-z_2)^2+(2C_{xx})^2}).
$$

While, of course, the parameter range for which physical density operators
compatible with those data exist is more limited in the former case,
it is indeed apparent that
the knowledge of $C_{xx}$ and $C_{zz}$ in combination with $z_1,z_2$ allows us
to infer a larger amount of entanglement.

This example highlights the importance of including all available
information in the entanglement verification as it may substantially
alter our conclusions.
The exact details of the procedure will, of course, depend on the concrete
situation.
%
\section{Multi-partite entanglement}
Our considerations are not restricted to bi-partite entanglement. Again,
quite general observables may be considered but in line with the bi-partite
case we illustrate this setting for a simple set of observables.
Let us consider the expectation values
$C_{xxx}=\langle \sigma_x\otimes\sigma_x\otimes\sigma_x\rangle$,
$C_{1zz}=\langle \id\otimes\sigma_z\otimes\sigma_z\rangle$ and
$C_{zz1}=\langle \sigma_z\otimes\sigma_z\otimes\id\rangle$. Given
the symmetries that leave these expectation values invariant, we
may restrict attention to density operators of the form
\footnote{Note that these states are diagonal in the GHZ-basis 
made up of the state $(|000\rangle\pm|111\rangle)/\sqrt{2}$,
$(|001\rangle\pm|110\rangle)/\sqrt{2}$,
$(|010\rangle\pm|101\rangle)/\sqrt{2}$ and
$(|011\rangle\pm|100\rangle)/\sqrt{2}$.}
\begin{equation}
        \sigma = \left(\begin{array}{cccccccc}
        a & 0 & 0 & 0 & 0 & 0 & 0 & h\\
        0 & b & 0 & 0 & 0 & 0 & g & 0\\
        0 & 0 & c & 0 & 0 & f & 0 & 0\\
        0 & 0 & 0 & d & e & 0 & 0 & 0\\
        0 & 0 & 0 & e & d & 0 & 0 & 0\\
        0 & 0 & f & 0 & 0 & c & 0 & 0\\
        0 & g & 0 & 0 & 0 & 0 & b & 0\\
        h & 0 & 0 & 0 & 0 & 0 & 0 & a
        \end{array}\right).
        \label{form}
\end{equation}
In the tripartite setting it is considerably more difficult than in the
bi-partite setting to define entanglement measures \cite{Plenio V 06}.
We consider two entanglement measures, the relative entropy of entanglement and the robustness of enanglement.

We begin with the relative entropy of entanglement with respect to Tri-PPT
states, i.e.\ states that are PPT with respect to any of the three possible
bi-partite cuts
\begin{equation}
        E_3(\sigma) = \inf_{\rho} \{S(\sigma||\rho): \rho\mbox{ is Tri-PPT}\}.
\end{equation}
It is helpful to note that it is always sufficient to restrict
the minimization over $\rho$ to those states that possess the same
local symmetries as $\sigma$ \cite{Vedral PRK 97,Vedral P 98}.
Thus only states $\rho$ of the form eq.\ (\ref{form}) need to be considered. These states all
commute with $\sigma$. Thus we are looking for a two-fold minimization
\begin{eqnarray}
        E_{\min} &=& \min_\sigma \{\min_\rho \{S(\sigma||\rho):\rho\mbox{ is Tri-PPT}\}:\\
                & &      \quad Tr[\sigma A_i]=a_i\}.\nonumber
\end{eqnarray}
We note that states $\sigma$ of the form eq.\ (\ref{form})
are Tri-PPT if and only if $\Delta= \max\{|e|,|f|,|g|,|h|\}-\min\{a,b,c,d\}\le
0$. Due to unitary invariance of the relative entropy we can apply local unitaries
to both $\rho$ and $\sigma$; one can therefore restrict to non-negative real $e$, $f$, $g$ and $h$.
Defining $m:=\min\{a,b,c,d\}$ we obtain the restrictions $e,f,g,h\le m$.

The expectation values for such states are given by
\beas
C_{xxx} &=& 2(e+f+g+h) \\
C_{1zz} &=& 2(a-b-c+d) \\
C_{zz1} &=& 2(a+b-c-d) \\
      1 &=& 2(a+b+c+d).
\eeas
Note that these expectation values lie in the range $[-1,1]$.

The minimisation over $\rho$ reduces to a three-parameter minimisation.
Let the matrix elements of $\sigma$ and $\rho$ (in the form (\ref{form})) be denoted
$a_\sigma$, $a_\rho$, etc. The three parameters are $a_\rho$, $b_\rho$ and $c_\rho$.
The other matrix elements of the optimal $\rho$ are given by
\beas
d_\rho &=& 1-(a_\rho+b_\rho+c_\rho) \\
h_\rho &=& \min(m_\rho,(h_\sigma/a_\sigma)a_\rho) \\
g_\rho &=& \min(m_\rho,(g_\sigma/b_\sigma)b_\rho) \\
f_\rho &=& \min(m_\rho,(f_\sigma/c_\sigma)c_\rho) \\
e_\rho &=& \min(m_\rho,(e_\sigma/d_\sigma)d_\rho),
\eeas
where $m_\rho:=\min\{a_\rho,b_\rho,c_\rho,d_\rho\}$.
The expression for the relative entropy in this optimal state is
\be
S(\sigma||\rho) = H((a_\sigma+h_\sigma,a_\sigma-h_\sigma)||(a_\rho+h_\rho,a_\rho-h_\rho))+\ldots
\ee
with three additional terms of obvious form.
Here, $H$ is the classical (Kullback-Leibler) relative entropy between two (unnormalised) two-dimensional
probability vectors.

\begin{figure}[tbh]
\centerline{
\includegraphics[width=8.5cm]{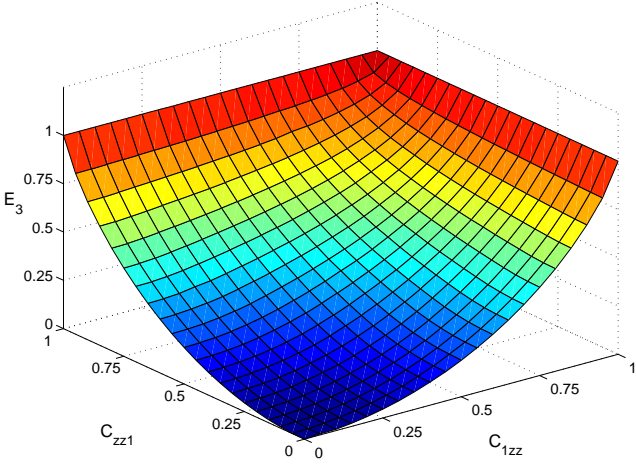}}
\caption{\label{relent}
For the given value of $C_{xxx}=1$, we plot the minimal
amount of entanglement, as measured by the relative entropy of entanglement $E_3$ w.r.t.\ Tri-PPT
states, consistent with the observation of $C_{1zz}$ and $C_{zz1}$.
}
\end{figure}

Because of joint convexity of the relative entropy, and convexity of the feasible set for $\rho$,
the remaining minimisation (over $\rho$ and $\sigma$)
is a convex one, which means that there can only be one local minimum. It can therefore be efficiently calculated
numerically using, e.g.\ conjugate gradient methods. We have performed numerical calculations
based on this method, and plotted the results in Figure \ref{relent} for the example of $C_{xxx}=1$.

Another possible entanglement quantifier is the random robustness
$R(\sigma)$ \cite{robustness}.
The random robustness is defined as the minimal amount of the maximally mixed state $\id/d$
that needs to be mixed with $\sigma$ to make the resulting state
Tri-PPT. Formally,
\be
R(\sigma) = \min_p \{p: p\id/\trace[\id] + (1-p)\sigma \mbox{ is Tri-PPT}\}.
\ee

We find that
\begin{equation}
        R(\sigma) = \max\left(0,\frac{\Delta}{1/8 +\Delta}\right).
\end{equation}
Therefore, the minimal robustness under the constraints
$C_{xxx}=\langle \sigma_x\otimes\sigma_x\otimes\sigma_x\rangle$,
$C_{1zz}=\langle \id\otimes\sigma_z\otimes\sigma_z\rangle$ and
$C_{zz1}=\langle \sigma_z\otimes\sigma_z\otimes\id\rangle$ is given
by
\begin{equation}
        R_{\min} = \max\left(0,\frac{\Delta_{\min}}{1/8 +\Delta_{\min} }\right),
\end{equation}
where
\begin{equation}
        \Delta_{\min} = \max\{\frac{|C_{xxx}| - 1}{2} + \frac{|C_{zz1}|+|C_{1zz}|}{4},0\}.
\end{equation}
%
\section{A general strategy for lower bounds on the negativity \label{LowerBound}}
In this Section, we readdress some of the issues of Section \ref{sec:purity}.
It is worth noting that the last result obtained there, eq.\ (\ref{eq27}),
could have been obtained from a general strategy to
obtain lower bounds for the minimization problem eq.\ (\ref{minimization}).

This can be achieved by using the fact that $||\rho^{\Gamma}||_1 =
\max_{||M||_{\infty}=1} \trace[M\rho^{\Gamma}]= \max_{||M||_{\infty}=1}
\trace[M^{\Gamma}\rho]$, where the maximization is over Hermitian $M$ \cite{Bhatia}.
Thus we consider the problem
\begin{eqnarray}
    E_{\min} &=& \log_2 \min_\rho \big\{ \max_M \{\trace[M^{\Gamma}\rho]:
    ||M||_\infty=1\} \nonumber\\
    &&\hspace*{2.cm}: \trace[\rho A_i]=a_i, F_i(\rho)\le f_i\big\},\;\;\;
\end{eqnarray}
where the outer minimization is over positive semidefinite matrices $\rho$
(the trace condition for states is included by putting $A_0=\id$, $a_0=1$),
and the inner maximisation is over all Hermitian matrices $M$.
When $F_i$ is a convex function its level sets $\{\rho: F_i(\rho)\le f_i\}$ are convex sets,
and we can use the minimax equality (see e.g.\ \cite{Boyd V 05}) to interchange inner and outer optimisations,
obtaining
\begin{eqnarray}
    E_{\min} \!\! &=& \!\!\log_2\! \max_M \big\{\! \min_\rho \{\trace[M^{\Gamma}\rho]: \trace[\rho A_i]=a_i, F_i(\rho)\le f_i \} \nonumber \\
    &&\hspace*{4.cm} : ||M||_\infty=1 \big\},
    \label{firstversion}
\end{eqnarray}
Let us now consider the case that there are no non-linear constraints $F_i$,
then the inner minimization is a semidefinite program (SDP).
We now apply Lagrange duality to this
minimization, i.e.\ we consider the unconstrained minimization of the Lagrangian
$\min_\rho \trace[(M^{\Gamma} - \sum \nu_i A_i)\rho]+\sum_i \nu_i a_i$ over all positive semidefinite $\rho\ge0$,
where the $\nu_i$ are the Lagrange multipliers.
If $M^{\Gamma} - \sum \nu_i A_i$ has negative eigenvalues, the minimum of the Lagrangian will be $-\infty$
(by letting $\rho$ become arbitrarily large),
and will not contribute to the outer maximization over $M$. Thus we can safely
require $M^{\Gamma} - \sum \nu_i A_i \ge 0$, in which case the minimum is obtained for $\rho=0$ and equals $\sum_i \nu_i a_i$.
Inserting this we find
\begin{eqnarray}
    E_{\min} &\ge& \log_2 \max_M \big\{\max_{\nu_i} \{\sum_i \nu_i a_i: \sum \nu_i A_i \le M^\Gamma \} \nonumber\\
    && \hspace*{3.cm}:||M||_{\infty}=1\big\}.
    \label{finalopt}
\end{eqnarray}
Because the inner minimization is an SDP, if the problem is strictly feasible, i.e.\ if all
inequality constraints can be satisfied with strict inequalities,
then we have strong duality \cite{Boyd V 05} and the above step
does not weaken the lower bounds.

Any choice of $M$ and $\nu_i$ such that $M^{\Gamma}\ge\sum_i\nu_i A_i$ and
$||M||_{\infty}=1$ now yields a lower bound on $E_{\min}$. Indeed, this could have been read off immediately from eq.\ (\ref{firstversion}).
However, as the optimization problem eq.\ (\ref{finalopt})
shows this may be overly restrictive. See \cite{Guehne RW 06,Eisert
BA 06} for lower bounds on other entanglement measures.

{\em Applications --} In the case of given $(C_{xx},C_{zz})$
with $C_{xx}+C_{zz}\ge 1$ as discussed in Section \ref{sec:xxzz},
we have $A_0=\id$, $A_1=\sigma_x\otimes\sigma_x$ and $A_2=\sigma_z\otimes\sigma_z$, and
$a_0=1$, $a_1=C_{xx}$ and $a_2=C_{zz}$. In this case we find as optimal $M$:
$$
M = \left(\begin{array}{cccc}
1&0&0&0 \\
0&0&1&0 \\
0&1&0&0 \\
0&0&0&1
\end{array}\right),
$$ (which indeed has operator norm 1)
and as optimal $\nu_i$:
$\nu_0=0,\nu_1=\nu_2=1$. One checks that $M^\Gamma\ge A_1+A_2$.
From this we recover again eq.\ (\ref{cxxczz}).

For the case of given
$(C_{xx},C_{yy},C_{zz})$ we choose $\nu_0=\nu_1=\nu_3=1/2, \nu_2=-1/2$,
so that $\sum_i \nu_i A_i = (\id +\sigma_x\otimes\sigma_x-\sigma_y\otimes\sigma_y+\sigma_z\otimes\sigma_z)/2$.
Taking $M=(\id +\sigma_x\otimes\sigma_x+\sigma_y\otimes\sigma_y+\sigma_z\otimes\sigma_z)/2$ (which has operator norm 1) yields
$M^\Gamma = \sum_i \nu_i A_i$,
and we recover the exact value found in eq.\ (\ref{eq27}).

{\em Proof of eq.\ (\ref{lowerpurity}) --}
A similar approach may suggest itself for the case concerning
purity and correlations discussed in section III and will be used to prove the lower bound eq.\ (\ref{lowerpurity}).
The constraints are however non-linear.
To proceed, we will use a kind of linearization procedure.
We begin by rewriting the quantities ${\cal C}_{zz}(\rho)$ and $\trace[\rho^2]$ in terms of expressions {\it
linear} in the tensor product $\tau:=\rho\otimes\rho$. Taking into
account $\trace[\rho]=1$ we find
\begin{eqnarray}
    {\cal C}_{zz}(\rho) &=& \trace[\tau Z]\;\;\;\mbox{and}\;\;\;
    \trace[\rho^2] = \trace[\tau F]
\end{eqnarray}
where $Z$ is the operator
\begin{displaymath}
    Z = \sigma_z\otimes\sigma_z\otimes\id\otimes\id -
    \id\otimes\sigma_z\otimes\sigma_z\otimes\id
\end{displaymath}
and $F$ is the flip operator that interchanges parties $1,2$ of the
first copy with parties $3,4$ of the second. The $Z$ presented here is the
simplest one that represents ${\cal C}_{zz}$. However, it is beneficial to use the symmetrised
form $Z'=(Z+FZF)/2$.

Let us now address the minimization
of $\frac{1}{2} \log_2 ||\rho^{\Gamma}\otimes \rho^{\Gamma}||_1$
given constraints on ${\cal C}_{zz}(\rho)$ and $\trace[\rho^2]$.
This problem is linear in $\sigma=\rho\otimes\rho$ and is therefore an
SDP. Consequentially, we can apply the above approach. Indeed, let us
choose $M = \diag(1 1 1 1\; 1 1 0 1 \; 1 0 1 1\; 1 1 1 1)$. Then, clearly,
$M^{\Gamma}-Z'/2-F\ge 0$ and we obtain eq.\ (\ref{lowerpurity}) as a lower bound on the
entanglement. This bound is certainly not tight, however. Indeed,
we could not have expected much more, as the extension of the problem to two copies allowed for much greater
freedom in the matrix $M$ and, therefore, led us
to underestimate the true value of $E_{\min}$.
\section{Verification of other physical properties}
In this work we have pointed out that in an experimental
verification of entanglement we need to search for the
least entangled state compatible with the measured data.
If the state so identified is entangled then the experimental
data prove the presence of entanglement. This approach
is not restricted to the verification of entanglement.
In fact, it applies to any physical property that we cannot
or chose not to measure directly.

Consider the property $\Pi$ of a quantum system which is
quantified by $\Pi(\rho)$.  If we are obtaining experimental
data, for example quantum mechanical averages of some
observables $A_i$, then we need to answer the

{\em {\bf\em Fundamental Question}: What is the least
value of $\Pi$ for which there is a state that is
compatible with the available measurement data?}

This smallest value of $\Pi$ is the value to which we
have verified the presence of $\Pi$. Mathematically
this may again be formulated as a minimization problem
in which the property $\Pi$ in the underlying quantum
state must be minimized subject to the positivity,
Hermiticity and normalization and measurement data
obtained as expectation values of observables $A_i$
or some non-linear function $F_i(\rho)$ of the density
matrix. Then the minimal amount of entanglement $E_{\min}$
under the given constraints is given by
\begin{eqnarray}
    \Pi_{\min} = \min_{\rho}\{
    \Pi(\rho): \trace[\rho A_i]=a_i, F_i(\rho)=f_i
    \}
    \label{minimization1}
\end{eqnarray}
where the minimisation domain is the set of states $\rho$.

In this more general framework the minimization of
entanglement is merely a special case of a general
approach to the verification of physical properties
in experiments.

\section{Summary and Conclusions}
In this work we have addressed the question of when
correlations or other measurement data that have been
observed in the classical measurement record of a
quantum system imply the existence of quantum correlations
in the underlying state. The {\em fundamental question}
in this area may be formulated as: {\em What is the
entanglement content of the least entangled quantum state
that is compatible with the available measurement data?}
We have formulated this question mathematically as an
optimization problem, discussed it for various examples
and provided some techniques for obtaining non-trivial
lower bounds on the minimal entanglement compatible with the
measurement data. The approach is equally valid in the
bi-partite and the multi-partite setting and for sub-systems
of arbitrary dimensionality. We hope that these investigations
will be helpful in experimental efforts that aim at the creation
and subsequent unequivocal verification and quantification of the
generated entanglement. This
should, in particular, apply to experimental set-ups where for
various reasons only a limited number of measurement settings is
available.

{\em Acknowledgements --}
We thank Alvaro Feito for careful reading of the manuscript and
helpful suggestions. We are grateful to Pawel Horodecki for
bringing ref.\ \cite{Horodecki HH 99} to our attention.
We also thank an anonymous referee for pointing out a simpler proof
of eq.\ (\ref{th1}) than the one contained in an earlier version.

This work was supported by The Leverhulme Trust, The Institute for 
Mathematical Sciences at Imperial College, the
Royal Society and is part of the QIP-IRC (www.qipirc.org)
supported by EPSRC (GR/S82176/0), the EU Integrated Project
Qubit Applications (QAP) funded by the IST directorate as
contract no.\ 015848.


\begin{thebibliography}{99}
\bibitem{footnote2} Note that, while formally
not dissimilar, this fundamental question is from a physical
point of view quite different from that concerning the maximization
under some constraints (see for example the concept of maximally
entangled mixed
states \cite{Ishizaka H 00,Verstraete AD 01}).
\bibitem{footnote3} It should be noted that the same lower bound
also holds true for the asymptotic relative entropy of entanglement \cite{Audenaert
EJPVD 01} as the lower bounds derived here are in fact additive.
\bibitem{footnote} For qubits the logarithmic negativity
\cite{Plenio 05,Plenio V 06} is the entanglement cost under exact
preparation using PPT-preserving operations \cite{Audenaert PE 03}
which in turn is larger than the entanglement cost under
PPT-preserving operations. This is lower bounded by the quantum
relative entropy of entanglement with respect to PPT-states. For
qubits, however, this is the same as the relative entropy with
respect to separable states, so that we have the same lower bound
on the logarithmic negativity in the plot.
\end{thebibliography}
\end{document}